\documentclass[aps,prl,twocolumn,showpacs,floatfix, superscriptaddress]{revtex4}
\usepackage{graphicx}
\usepackage{amsfonts}
\usepackage{amsmath}
\usepackage{bm}

\begin{document}

\title{Non-continuous Froude number scaling for the closure depth
of a cylindrical cavity}

\author{Stephan Gekle}
\author{Arjan van der Bos}
\author{Raymond Bergmann}
\author{Devaraj van der Meer}
\author{Detlef Lohse}

\affiliation{Physics of Fluids Group and J.M. Burgers Centre for Fluid Dynamics, University of Twente, P.O. Box 217, 7500AE Enschede, The Netherlands}
\date{\today}

\begin{abstract}

A long, smooth cylinder is dragged through a water surface to create a cavity
with an initially cylindrical shape. This surface void then collapses due to the
hydrostatic pressure, leading to a rapid and axisymmetric pinch-off in a single
point. Surprisingly, the depth at which this pinch-off takes place does not
follow the expected Froude$^{1/3}$ power-law. Instead, it displays two
distinct scaling regimes separated by discrete jumps, both in experiment and in
numerical simulations (employing a boundary integral code). We quantitatively
explain the above behavior as a capillary waves effect. These waves are created when the top of the cylinder passes the water surface. Our work thus gives further evidence for the non-universality of the void collapse.

\end{abstract}

\pacs{47.55.D-,47.55.db,47.11.Hj,47.35.Pq}

\maketitle


Many phenomena in fluid dynamics are known to be universal, allowing physicists to describe their final outcome without precise knowledge of the initial conditions. One prime example for such universality is the pinch-off of a liquid droplet in air \cite{Brenner_PRL_droplet_fission, Eggers_RMP, Nagel_Science}.
The inverse problem, however, has recently turned out to be non-universal: When an air bubble pinches off inside a liquid, the local dynamics around the pinch-off point retains a memory of its creation until the very end \cite{PRL_disk,Gordillo_non_universal_JFM, Nagel_breakdown}. In this Letter we show that also the \textit{global}  dynamics of this problem is characterized by the absence of universality. To this end we examine the air-filled cavity created when a solid object is rapidly submerged through a water surface. The walls of the cavity subsequently collapse due to hydrostatic pressure from the liquid bulk. When the colliding walls meet, a violent jet shoots up into the air. 
Our experimental and numerical evidence illustrates how universality is broken through the interference of a second phenomenon unrelated to hydrostatic pressure. Surface waves created at the very beginning as the object passes the water surface exert a decisive influence until the end of the process. This leads to a discontinuous dependence of the pinch-off location on the velocity of the submerging object.


In our experiment we drag a cylinder with radius $R_0=20$ mm and length $l=147$ mm through the surface of a large water tank using a linear motor connected to the cylinder bottom by a rod. With this setup we prescribe the cylinder velocity $V$ between 0.5 and 2.5 m/s keeping it constant throughout the whole process. The relevant dimensionless parameter is the Froude number $\mathrm{Fr} = V^2/(R_0g)$ with $g=9.81$ $\mathrm{m/s^2}$, which in our experiment ranges between 1.2 and 32.

\begin{figure*}
\begin{center}
\includegraphics[width=3.5cm]{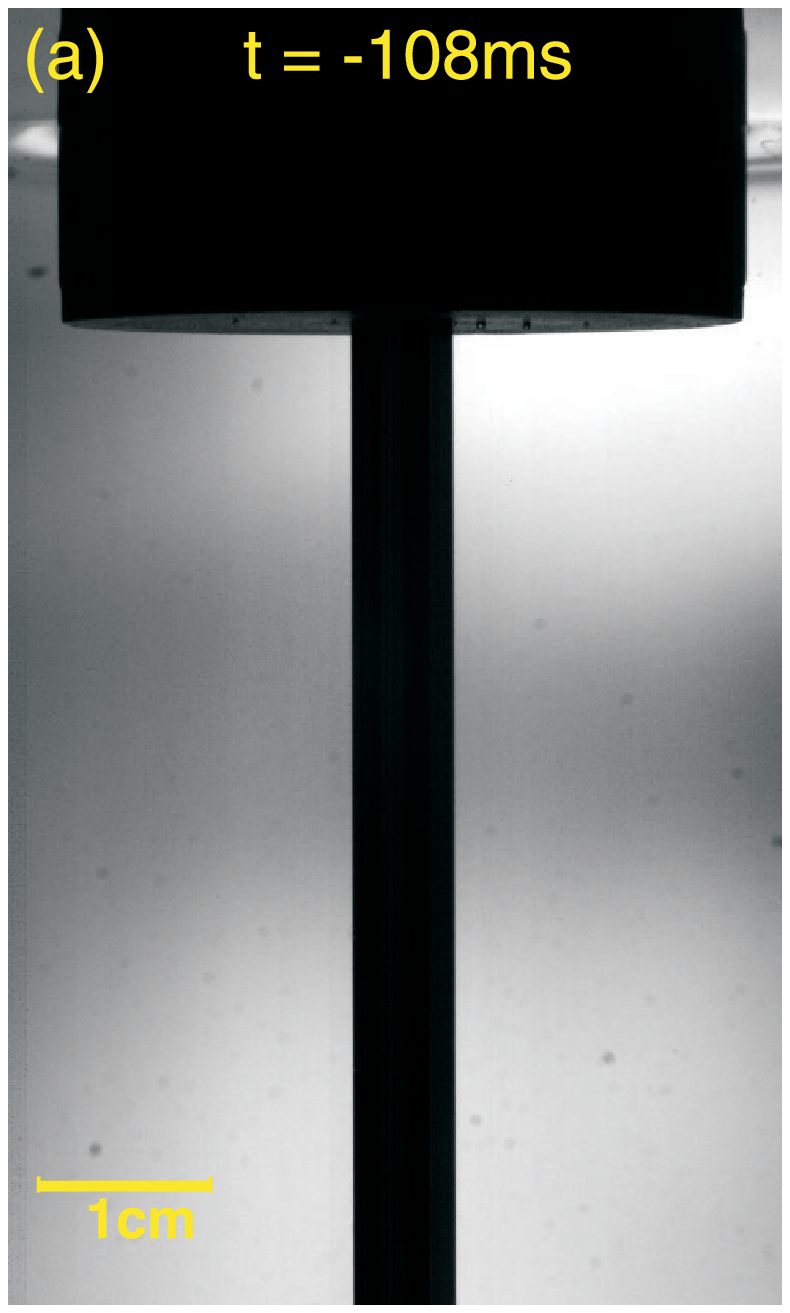}
\hspace{0.3cm}
\includegraphics[width=3.5cm]{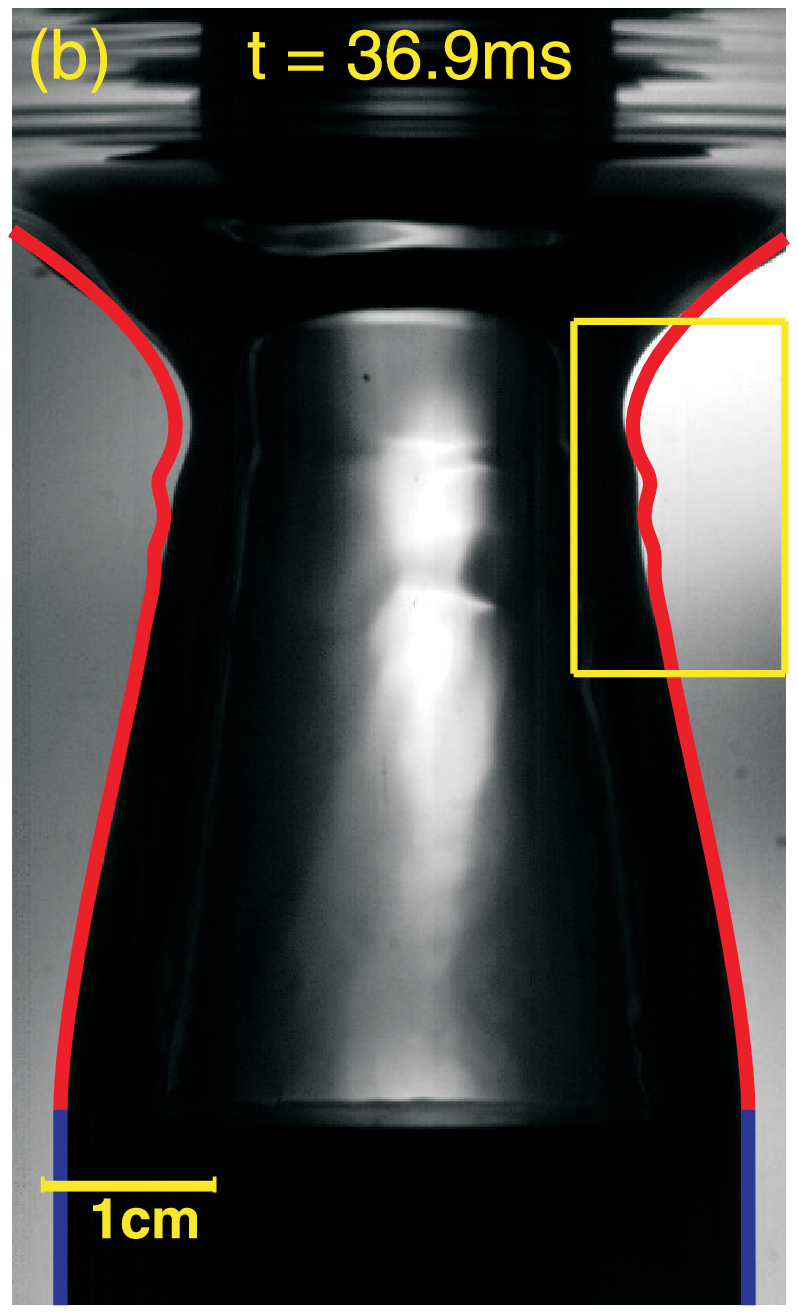}
\hspace{0.3cm}
\includegraphics[width=3.5cm]{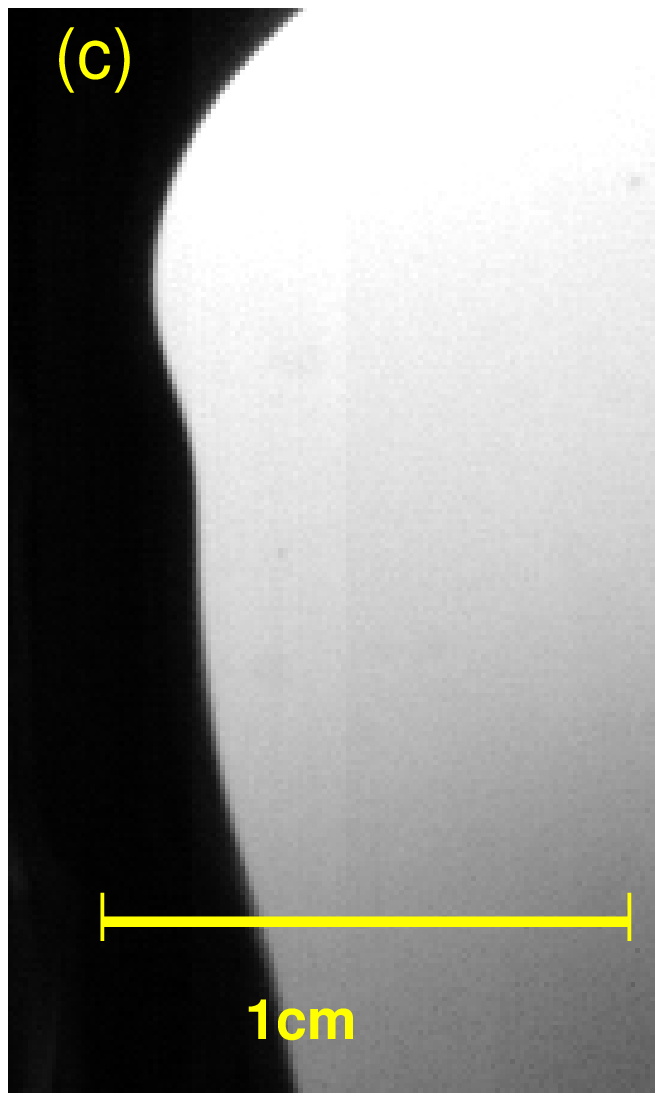}
\hspace{0.3cm}
\includegraphics[width=3.5cm]{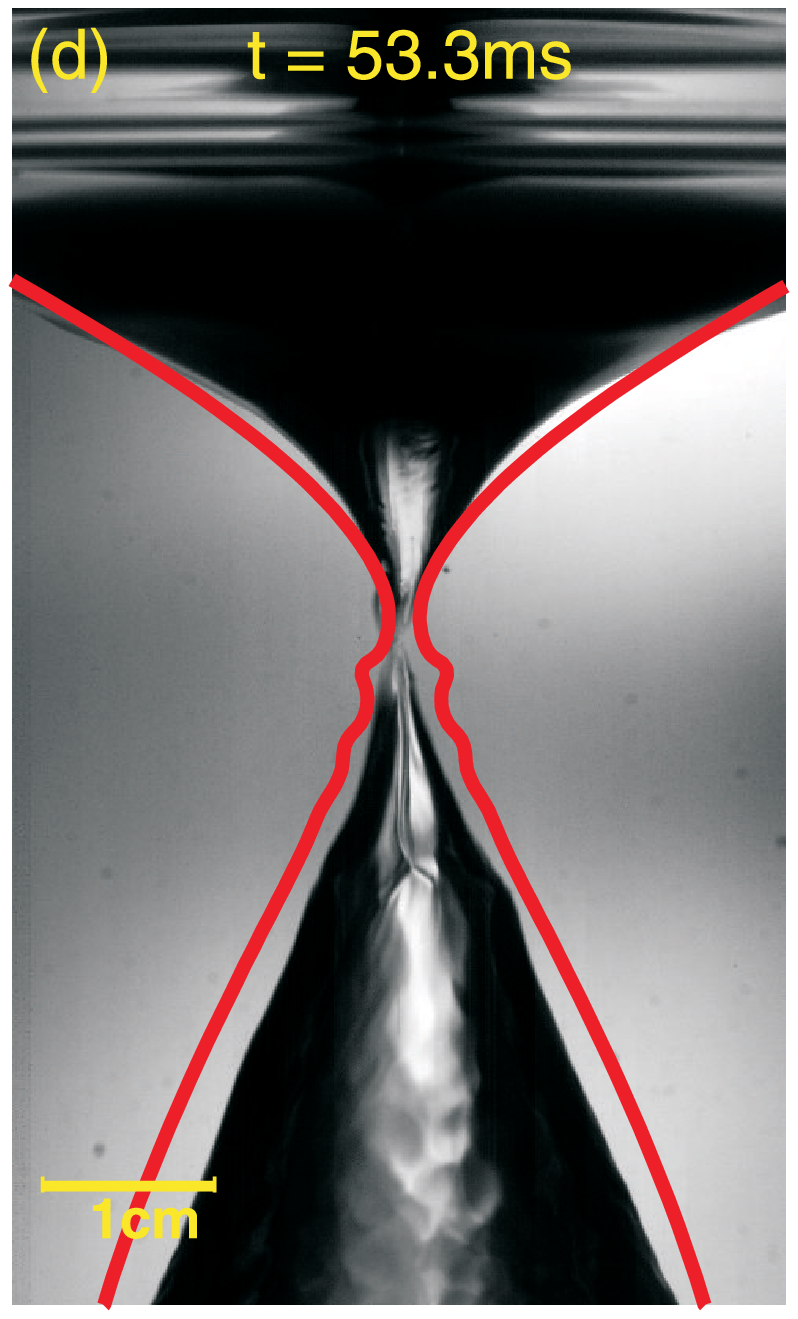}
\end{center}
\caption{(color online) Snapshots of the cavity created by the submerging cylinder from the starting position (a) until pinch-off (d) taken at a cylinder speed of $V=1.5$ m/s ($\mathrm{Fr}=11.5$). The area designated by the yellow rectangle in (b) is magnified in (c) illustrating the rippled surface of the cavity walls. Red and blue lines show the free surface and the cylinder, respectively, taken from the numerical simulation. In (a) and (b) video frames are connected to simulation time by matching the cylinder position. Due to the fact that the precise closure time sensitively depends on the amplitude of the capillary waves traveling over the cavity surface, for (d) it was more convenient to match the cavity closure times instead.}
\label{fig:sequence}
\end{figure*}

The shape of the axisymmetric cavity is imaged with a high speed camera at up to 10,000 frames/sec, with the vertical coordinate $z$ pointing upwards along the cylinder axis and $r$ being the radial coordinate. Figure~\ref{fig:sequence} shows a typical sequence of the cavity dynamics. We choose the starting position of the lower edge of the cylinder slightly below the water surface to suppress the splash (a). From there it is pulled downwards with high acceleration such that it has reached its prescribed speed before the top passes the water surface at $t=0$. The submerging cylinder creates an air-filled cavity whose side walls immediately start to collapse (b). Magnification of the cavity walls as in (c) shows that they are not smooth surfaces, but exhibit pronounced ripples. The moment of pinch-off is shown in (d).
 
\begin{figure}
\includegraphics[angle = 270, width=\columnwidth]{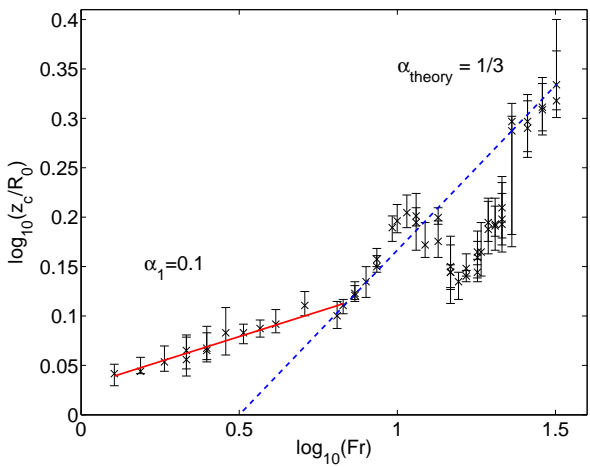}
\caption{(color online) The experimental closure depth as a function of the Froude number. The asymptotic regime for low Froude numbers scales with an exponent $\alpha_1\approx 0.1$ (red solid line). Only in the limit of high Froude numbers is the data consistent with an exponent of $\frac{1}{3}$ (blue dashed line) as in \cite{Clanet_JFM, Andrea_liquid_cylinder, Sandjet_PRL}. For the intermediate regime the experiments do not allow the extraction of a systematic behavior.}
\label{fig:Froude_scaling_exp}
\end{figure}

From the high-speed video images we extract the closure depth of the cavity $z_c$ for cylinder velocities up to 2.5 m/s. For higher velocities the collapse becomes so violent that the video images do not allow for a clear determination of the closure depth, thus limiting the range of our experiments. Plotting the closure depth over the Froude number as in Fig.~\ref{fig:Froude_scaling_exp}, we find two asymptotic regimes. The regime for low Froude numbers obeys a scaling behavior $z_c\sim\mathrm{Fr}^\alpha$ with an exponent $\alpha=\alpha_1\approx 0.1$. This is in contrast to earlier experiments on the impact of similar cylindrical objects \cite{Clanet_JFM, Andrea_liquid_cylinder}. These showed a single continuous scaling behavior with $\alpha\approx\frac{1}{3}$, for which a theoretical explanation could be given \cite{Clanet_JFM, Andrea_liquid_cylinder, Sandjet_PRL}. Only for high Froude numbers is our data consistent with this $\frac{1}{3}$-scaling law. In between these two regimes we find a transition region without any experimentally discernable structure.

To understand the underlying mechanism leading to this discontinuous behavior, we have conducted boundary integral simulations. This method \cite{Andrea_BI} assumes an inviscous and vorticity-free fluid. In fact, with the kinematic viscosity $\nu$ the global Reynolds number $\mathrm{Re} = R_0V/\nu$ is of the order of $10^4$, while the local Reynolds number $\mathrm{Re} = R\dot{R}/\nu$ defined with the cavity radius $R$ for the point of minimum radius lies between $10^2$ and $10^5$. We can thus safely neglect viscosity. Further, due to the small density of air as compared to water, its dynamic effects can be neglected for the study of the global cavity shape. Our numerical results can be compared to the high-speed videos without the use of any adjustable parameter. They are in good agreement with experimental observations as demonstrated by the colored lines in Fig.~\ref{fig:sequence}.

These numerical simulations allow us to study the ripples observed in Fig.~\ref{fig:sequence} (c) in great detail. As the cylinder top passes the water surface, the rectangular corner between the cylinder wall and the water surface is no longer held in place by the solid boundary of the cylinder. The newly created free surface thus possesses a corner with very high (initially infinite) curvature. Surface tension immediately tries to flatten this surface by pulling the corner diagonally inwards into the fluid bulk. This results in a shock similar to throwing a stone onto a lake, which consequently leads to the formation of capillary waves travelling out- and downward on the free surface. The downward waves can be observed in Fig.~\ref{fig:sequence} (c). Such a shock creates a wave packet containing waves of different frequencies. Each of these waves spreads with a velocity $c=\omega/k$ given by the dispersion relation $\omega^2=\left(\sigma/\rho\right)k^3$ of capillary waves, where $\omega$ is the angular frequency, $k$ the wave vector, $\sigma$ the surface tension coefficient and $\rho$ the density. Using the Fourier transform of this initially localized packet, we can apply stationary phase approximation and thus calculate the dominant wave vector $k^*$ at a given distance $x$ from the source at time $t$:
\begin{equation}
k^* = \left(\frac{2}{3}\frac{x}{t}\right)^2\frac{\rho}{\sigma}. \label{eqn:waves}
\end{equation} 

\begin{figure*}
\hspace{-1cm}
\includegraphics[height=5cm, width=0.6\columnwidth]{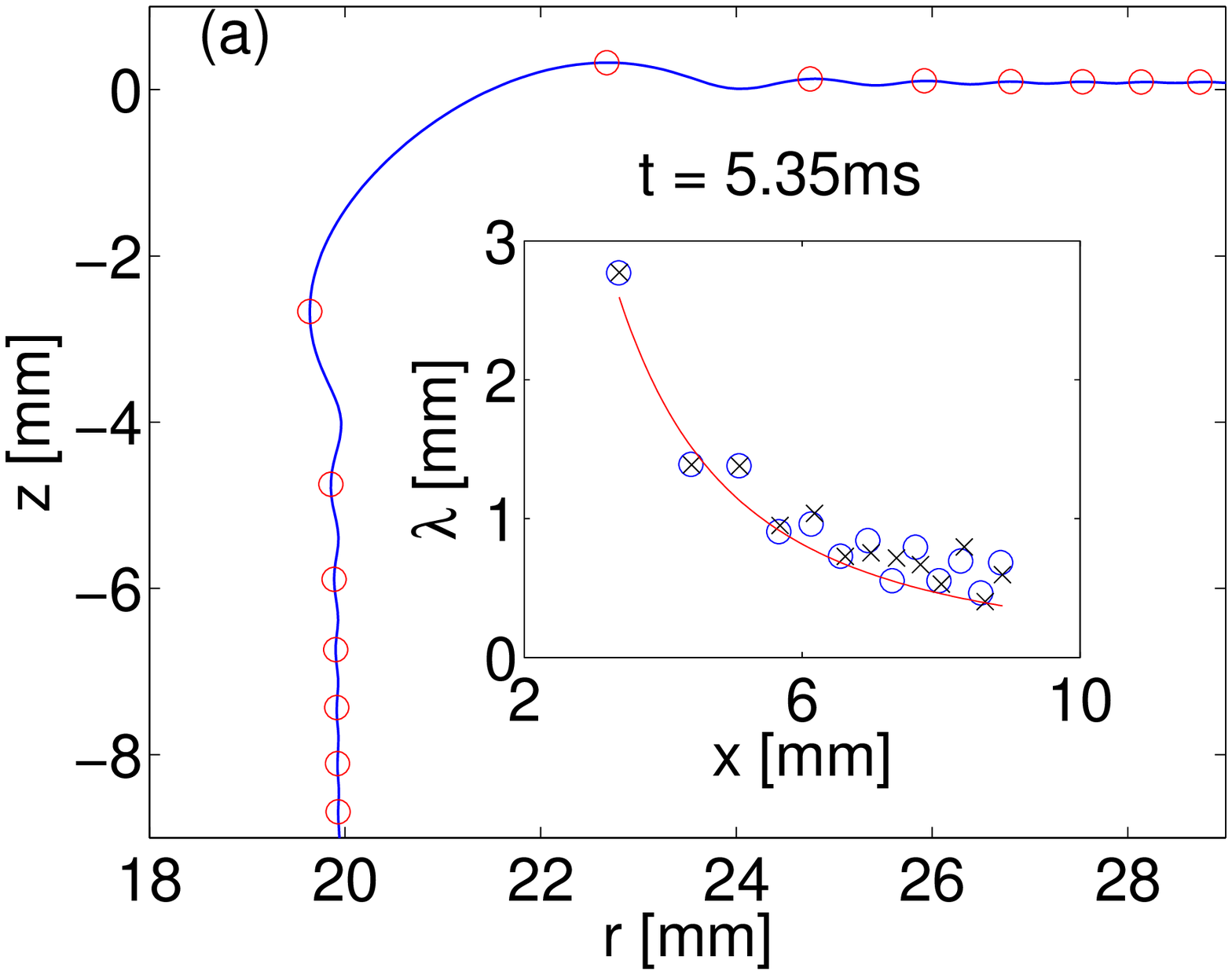}
\includegraphics[height=5cm, width=1.4\columnwidth]{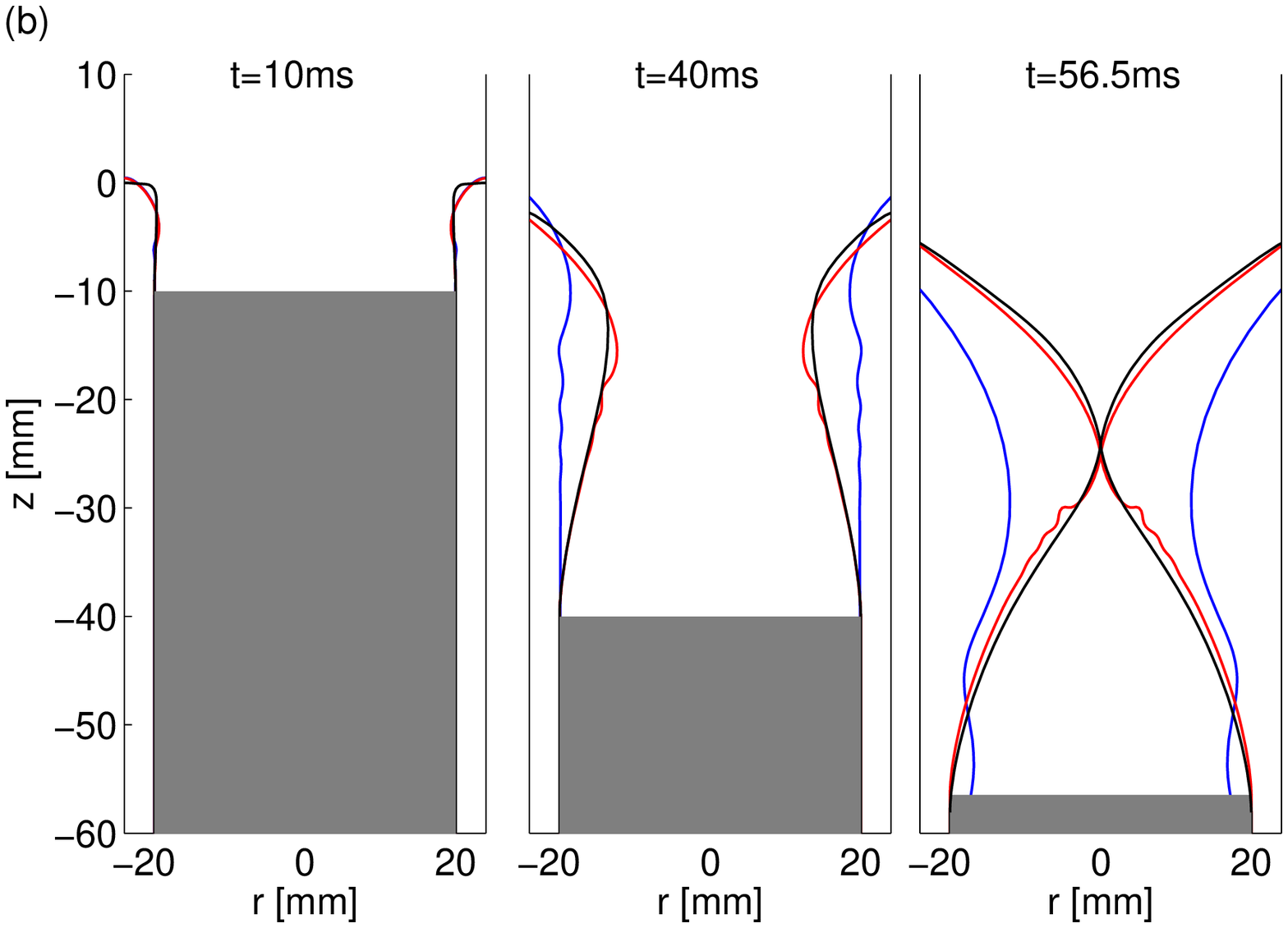}
\caption{(color online) (a) Capillary waves at $t=5.35$ ms after the cylinder top has passed the water surface with $V=3$ m/s ($\mathrm{Fr}=45.9$) as obtained from the simulation. Wave crests are marked by red circles and allow an estimate of the dominant wave length $\lambda=2\pi/k$ at a given position. The inset compares these wave lengths for the downward (blue circles) and outward (black crosses) waves to the theoretically expected behavior for capillary waves of Eq.~(\ref{eqn:waves}) (red line). 
(b) The natural cavity dynamics (red) as a superposition of two hypothetical settings: without surface tension (black) and without gravity (blue) for $V=1$ m/s ($\mathrm{Fr}=5.1$).}
\label{fig:waves_hypo}
\end{figure*}

Our simulations are detailed enough to allow for an accurate estimate of the dominant wave length $\lambda$ from the shape of the surface, cf. Fig.~\ref{fig:waves_hypo} (a). Comparison with Eq.~(\ref{eqn:waves}) as in the inset of Fig.~\ref{fig:waves_hypo} (a) confirms that the observed ripples on the surface are indeed capillary waves originating from the corner point as the cylinder top passes the water surface.

The closing of the cavity is driven by hydrostatic pressure which acts on every point of the vertical free surface accelerating it inward as soon as the cylinder has passed. This accelerating force increases with the depth. Thus, points near the top surface start moving early with a small acceleration, while deeper points start with increasing delay, but higher acceleration \cite{Clanet_JFM,Sandjet_PRL}. On a rippled surface this process favors the wave crests over the other points. The resulting closure depth will thus be determined by a combination of (i) hydrostatic pressure induced by gravity and (ii) capillary waves created by surface tension. 

\begin{figure}
\includegraphics[angle = 270, width=\columnwidth]{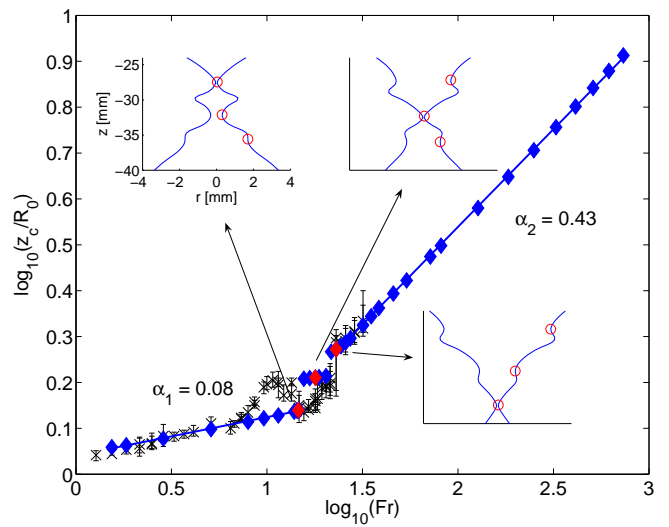}
\caption{(color online) Comparison of the experimental closure depth (black crosses) with the numerical data (blue diamonds). The dependence of the closure depth on the Froude number exhibits three clearly distinct regimes. The insets illustrate the shape of the cavity at pinch-off for one representative of each regime (axis are the same for all insets). The regimes are determined by which local minimum first meets its counterpart on the central axis. }
\label{fig:Froude_scaling_sim_exp}
\end{figure}

The numerically obtained closure depth shown in Fig.~\ref{fig:Froude_scaling_sim_exp} convincingly reproduces the experimental results, while for very high Froude numbers only numerical data is available. The first regime shows a scaling behavior with $\alpha_1\approx0.08$, sufficiently close to the experimental value. The second regime scales with $\alpha_2\approx0.43$, which is somewhat off the expected value of $\frac{1}{3}$ from \cite{Clanet_JFM, Andrea_liquid_cylinder, Sandjet_PRL}, but well compatible with our experimental observations. The simulations even make the identification of an intermediate regime possible, which due to its small range in Froude numbers cannot be clearly observed in experiments. The slight shift between numerical and experimental data on the Froude axis can be attributed to small contaminations which lower the surface tension of the water.

The insets in Fig.~\ref{fig:Froude_scaling_sim_exp} elucidate precisely how the capillary waves lead to the discontinuous jumps between the different regimes of the closure depth: For Froude numbers near the regime transition three different local minima of the radius come very close to meeting their counterpart on the opposite side. In the first regime the uppermost of the three minima closes first and thus determines the closure depth, while in the second regime the one located in the middle is the fastest to reach the central axis. Finally, when the lowest minimum closes before the other two the third regime is attained.

\begin{figure}
\includegraphics[angle=270, width=\columnwidth]{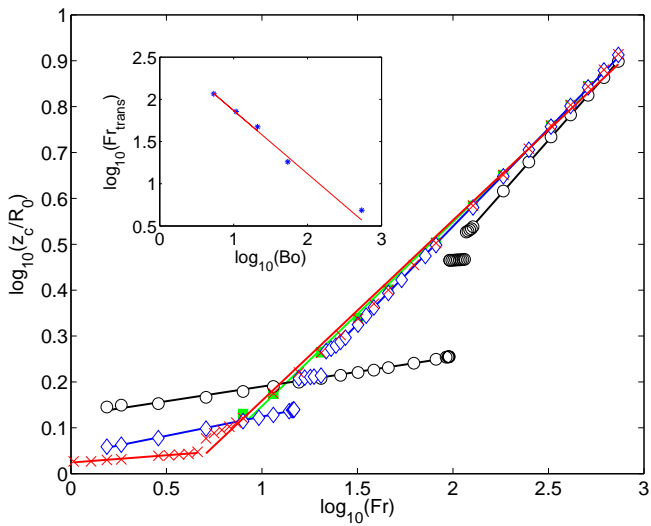}
\caption{(color online) The numerical closure depth as a function of the Froude number for different surface tension coefficients: $\sigma_1 = 728$ $\mathrm{mN/m}$ (black circles), $\sigma_2=72.8$ $\mathrm{mN/m}$ (blue diamonds), $\sigma_3 = 7.28$ $\mathrm{mN/m}$ (red crosses) and  $\sigma_4=0$ (green squares). A higher $\sigma$ leads to a significant extension of the first regime. The inset shows the onset of the high Froude number regime as a function of the Bond number with the red line depicting the expected scaling law $\mathrm{Fr_{trans}}\sim\mathrm{Bo}^{-3/4}$.}
\label{fig:Froude_scaling_sigma}
\end{figure}

Since the behavior of capillary waves is determined by surface tension, one expects that modifications of the surface tension coefficient $\sigma$ should significantly alter the closure depth in the first and any intermediate regimes. For the last regime capillary waves are irrelevant and the scaling behavior of the closure depth can be derived independent of surface tension \cite{Clanet_JFM, Andrea_liquid_cylinder, Sandjet_PRL}. In numerical simulations a significant change of the surface tension coefficient is feasible and we have performed simulations with a ten-fold increase and decrease of $\sigma$ as compared to the natural value of 72.8mN/m for water as well as for a hypothetical liquid without any surface tension ($\sigma=0$). As Fig.~\ref{fig:Froude_scaling_sigma} demonstrates, the Froude number ranges for the three regimes are found to depend indeed strongly on the value of the surface tension coefficient. The length of the first regime significantly enlargens for a higher surface tension coefficient. As expected, the last regime is almost uninfluenced by changes in surface tension. For low surface tension merely the two asymptotic regimes exist and the intermediate regime is not observed anymore. The limiting case completely without surface tension possesses only one regime, consistent with the $\mathrm{Fr}^{1/3}$-scaling.

Using Eq.~(\ref{eqn:waves}) with $x=z_c$ and $k^*\sim1/z_c$ we can estimate the relevant time scale for the capillary waves as $t_w\sim\sqrt{\rho/\sigma}z_c^{3/2}$ while the relevant time scale for cavity closure is $t_c\sim z_c/V$. The onset of the high Froude number regime is now readily found by equating the ratio $t_w/t_c$ to a constant of order 1. Introducing the Bond number $\mathrm{Bo}=gR_0^2\rho/\sigma$ and making use of the theoretically expected scaling $z_c\sim\mathrm{Fr}^{1/3}$ this yields $\mathrm{Fr_{trans}}\sim\mathrm{Bo}^{-3/4}$ which is in good agreement with our observations as shown in the inset of Fig.~\ref{fig:Froude_scaling_sigma}.  

The cavity dynamics completely \textit{without surface tension} and thus deprived of all capillary waves is illustrated by the black lines in Fig.~\ref{fig:waves_hypo} (b). The blue lines depict the evolution of a free surface starting with a (nearly) rectangular corner \textit{without gravity} allowing us to study the formation of capillary waves in an isolated setting. The cavity dynamics under realistic conditions ($g=9.81$ $\mathrm{m/s^2}$ and $\sigma = 72.8$ mN/m) is shown in red and can clearly be identified as a superposition of the above mentioned limiting cases. In the first instants, the real dynamics is almost identical to the one without gravity with its main feature being the capillary waves. Later in the process hydrostatic pressure becomes more important until finally the cavity approaches the shape of the pure gravity simulation with the capillary waves superposed on the cavity walls.

Finally, to examine the local dynamics around the pinch-off point we extract from the simulation the neck radius and the radius of curvature at closure depth \cite{PRL_disk, Taborek_PRL}. Both scale with the time $\tau = t_c-t$ until pinch-off with scaling exponents $\alpha_{\mathrm{neck}}=0.54\ldots0.57$ and $\alpha_\mathrm{curvature}=0.28\ldots0.43$, which agrees well with the observations for an impacting disk in \cite{PRL_disk}. The diffferent scaling behavior for the two length scales suggests that the local dynamics exhibits similar non-universal features as in \cite{PRL_disk}.


In conclusion, we have shown that capillary waves created when a submerging object passes the water surface have a strong and lasting influence on the global dynamics of the cavity. This influence remains observable until the very end of the cavity collapse manifesting itself in clearly distinct regimes of the closure depth as a function of the submerging velocity. We have thus presented evidence that the pinch-off of air in water - in contrast to the pinch-off of a water droplet in air - is non-universal not only in a local \cite{PRL_disk, Nagel_breakdown}, but also in a global sense. Since capillary waves are an unavoidable consequence of disturbances on a water surface, we expect that the effects elaborated in this Letter will be of relevance to a wide range of related phenomena.

\begin{acknowledgments}
We thank L.\ van Wijngaarden 
and A.\ Prosperetti for stimulating discussions. This work is part of the research program of the Stichting FOM, which is financially supported by NWO.
\end{acknowledgments}

\end{document}